%% file: paper.tex
\documentclass[10pt, conference, compsocconf]{IEEEtran}
\pdfoutput=1
\ifCLASSINFOpdf
  \usepackage[pdftex]{graphicx}
  \DeclareGraphicsExtensions{.pdf,.jpeg,.png}
\else
  \usepackage[dvips]{graphicx}
  \DeclareGraphicsExtensions{.eps}
\fi

\usepackage{color}
\usepackage[cmex10]{amsmath}
\usepackage{graphics,latexsym,amssymb,xspace,enumerate,paralist}
\usepackage{algorithm}
\usepackage[noend]{algorithmic}
\usepackage{array}
\usepackage[tight,footnotesize]{subfigure}

\newcommand {\naren}[1]{}
\newcommand {\deb}[1]{}
\newcommand {\yong}[1]{}
\newcommand {\sean}[1]{}

\newtheorem{definition}{Definition}

\newcommand{\AOAT}{Accelerator-Oriented Algorithm Transformation }

\begin{document}
\title{Accelerator-Oriented Algorithm Transformation for Temporal Data Mining}


\author{\IEEEauthorblockN{Debprakash Patnaik, Sean P. Ponce, Yong Cao, Naren Ramakrishnan}
\IEEEauthorblockA{Department of Computer Science, Virginia Tech, Blacksburg, VA 24061, USA\\
Email: \{patnaik, ponce, yongcao, naren\} @vt.edu}
}

\maketitle

\begin{abstract}
\input{tex/abstract}
\end{abstract}


%\begin{CRcatlist}
%	\CRcat{H.2.8}{Information Systems}{Database Management}
%	{Database Applications}
%\end{CRcatlist}

\IEEEpeerreviewmaketitle

\input{tex/introduction}

\input{tex/background}

\input{tex/priorwork}
\input{tex/algorithms}

\input{tex/results}

\input{tex/discussion}

%\input{tex/repeatability}

%\section*{Acknowledgment}
%more thanks here

%\bibliographystyle{IEEEtran}
%\bibliography{bib/references}

\end{document}

%% file: tex/abstract.tex
Temporal data mining algorithms are becoming increasingly important
in many application domains including computational neuroscience, especially 
the analysis of spike train data. While application scientists have 
been able to readily gather multi-neuronal datasets, analysis capabilities 
have lagged behind, due to both lack of powerful algorithms and 
inaccessibility to powerful hardware platforms. The advent of GPU
architectures such as Nvidia's GTX 280 offers a cost-effective
option to bring these capabilities to the neuroscientist's desktop. Rather
than port existing algorithms onto this architecture, we advocate the need
for {\it algorithm transformation}, i.e., rethinking the design of the
algorithm in a way that need not necessarily mirror its serial 
implementation strictly. We present a novel implementation of a frequent
episode discovery algorithm by revisiting ``in-the-large'' issues
such as problem decomposition as well as ``in-the-small'' issues such
as data layouts and memory access patterns. This is non-trivial because
frequent episode discovery does not lend itself to GPU-friendly  
data-parallel mapping strategies. Applications to many datasets and
comparisons to CPU as well as prior GPU implementations
showcase the advantages of our approach.

%% file: tex/introduction.tex
\section{Introduction}

\begin{figure}[ht]
\centering
\includegraphics[width=0.75\columnwidth]{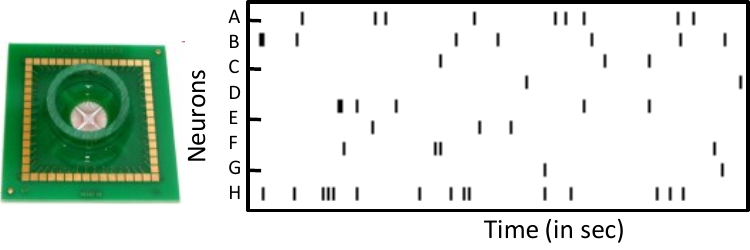}%
\caption{Illustration of a Micro electrode array (MEA) used to record
spiking activity of neurons in tissue cultures and the spike train recorded by the MEA.}%
\label{fig:mea}%
\end{figure}

\yong{Briefly talk about temporal data mining and its application in neuroscience. (one paragraph)}
Discovering frequently repeating patterns in event sequences is an important data mining problem that finds application in domains such as 
industrial plants/assembly lines, medical diagnostics, and
computational neuroscience. Algorithms such as frequent episode discovery~\cite{vatsan1,Patnaik2008}, in particular, are adept at discovering 
patterns in neuronal spike train data using multi-electrode arrays
(MEAs; shown in Fig.~\ref{fig:mea}).
%The goal of this data mining task is to find frequently repeating 
%local patterns amongst a 
%long sequence of discrete instantaneous events. In particular, the
%frequent episodes discovery framework~\cite{vatsan1} has found 
%useful application
%in the area of spike train analysis, e.g., see~\cite{Patnaik2008}. The 
%algorithms proposed in~\cite{Patnaik2008} are well suited for analyzing simultaneously recorded spike trains from living neuron cells using  multi-electrode arrays (MEA) (shown in Fig.~\ref{fig:mea}). The patterns discovered help 
%identify cascades of firing neurons and their characteristic delays. 
They bring us one step closer to reverse-engineering the temporal connectivity map of neuronal circuits and yield insights into the 
network level activity of brain tissue. Until recently, most analysis was limited to single cell activity due to the
low computational capacity of single-core computers and lack of efficient data mining algorithms. While clever data mining algorithms have made it 
feasible to overcome the combinatorial explosion problem of discovering 
multi-way interactions, they are still limited by the
memory and processing power constraints of the single CPU.

\yong{Mention about the computational requirement for mining large dataset, and introduce commoditized many-core GPGPUs (massive parallel and personal supercomputer)}

In recent years, the peak-performance of a GPU has exceeded that of
the CPU by several orders of magnitude. Intel's latest quad-core processors have a theoretical peak of 51.20 GFLOPS. Nvidia's latest single GPU card, the GeForce GTX 285, has a theoretical peak of 1062.72 GFLOPS. Application speedups up to 431x have been reported~\cite{Ryoo2008}. The demand for graphics applications 
like gaming have made GPUs widely available and inexpensive. In this paper we explore how this massively parallel computing platform can be used effectively to solve the challenges posed by temporal data mining.
Although we focus on a specific algorithm, the issues encountered here
are symptomatic of many temporal mining algorithms that must use state machines
to monitor and process event streams. Hence the lessons learned from this
effort will likely seed similar research efforts.

%\subsection{Motivation}

 \yong{The motivation:\\
    * The nature of original temporal data mining algorithm (lot of dependencies, non-data parallel) limits the performance gain of standard hardware-oriented optimization.}

One major challenge in utilizing the GPU lies in transforming algorithms to 
operate efficiently on the GPU. Another challenge is understanding how to use the GPU architecture to achieve maximum performance. Data-parallel algorithms require relatively less work to map computation onto the GPU architecture. For algorithms with complex data dependencies, such as dynamic programming, it is difficult to achieve a large speedup. The nature 
of the temporal data mining problem addressed in this work limits the 
performance gain of standard hardware-oriented optimizations applied to 
direct ports of existing algorithms (due to their sequential dependencies
implicit in processing event streams).

\yong{Main contribution:``In this project, we adopt the concept of AOAT and introduce a new algorithm for temporal data mining on GPUs.''}

In this paper, we adopt the concept of \AOAT and introduce a new algorithm for counting occurrences of frequent episodes with temporal constraints (a key
analysis task for event stream analysis)
on the GPU. Here an episode is a sequential dependency of the form `event 
A followed by event B followed by event C...' where
there could be don't care or ``junk'' events interspersed between
the pattern events. (The frequency of
such episodes is defined as the maximum number of non-overlapped occurrences
of the episodes in the event stream.)

The traditional approach to parallelizing existing algorithms is to start with the existing sequential algorithm, identify the data dependency patterns, and restructure the original algorithm to achieve maximum data parallelism. Another way is to look at the problem being solved and formulate a decomposition using known parallel primitives. We follow the later approach and design a new algorithm by decomposing the original problem into two sub-problems: finding overlapped episode occurrences and resolving overlaps to obtain non-overlapped counts. We introduce a parallel local tracking algorithm to solve the first sub-problem which is computationally more demanding. On the other hand,
the second sub-problem contributes only a very small overhead to the 
overall computation and is hence solved sequentially. The re-designed algorithm exhibits a higher degree of parallelism resulting in a performance gain over the sequential algorithm implemented on a single-core CPU and a
previously optimized GPU implementation (\textit{MapConcat}) \cite{Yong2009} which achieves parallelism by mining segments of the data sequence.

\yong{Briefly describe the concept of AOAT (based on my previous description).\\
We redesign the algorithm by decomposing the original problem into two sub-problems: overlapped episode tracking and overlap removal. We introduce a parallel local tracking algorithm to solve the first sub-problem, while the second sub-problem only take a very small portion of the overall computation. The re-designed algorithm exhibit a high degree of parallelism which produce substantial performance gain over single-core CPU implementation and an optimized GPU implementation of the original algorithm (MapConcat).}

%The rest of paper is organized as follows. 

%% file: tex/background.tex
\section{Background}
%# GPGPU
%
%    * GPU architecture and programming model
%    * Application optimization with GPUs. Mention that most of researches focused on computation-to-core mapping and hardware-based optimization.

\subsection{GPGPU Architecture}
To understand the algorithmic details behind our approach, we 
provide a brief overview of GPGPU and its architecture.

The initial purpose of specialized GPUs was to accelerate the display
of 2D and 3D graphics, much in the same way that the FPU focused on
accelerating floating-point instructions.  However, the rapid
technological advances of the GPU, coupled with extraordinary
speed-ups of application ``toy'' benchmarks on the specialized GPUs,
led GPU vendors to transform the GPU from a specialized processor to a
general-purpose graphics processing unit (GPGPU), such as the NVIDIA GTX
280.
%, as shown in Figure~\ref{fig:gpu}.  
To lessen the steep learning
curve, GPU vendors also introduced programming environments, such as
the Compute Unified Device Architecture (CUDA).

\begin{figure}[htbp]
\centering
\includegraphics[width=0.8\columnwidth]{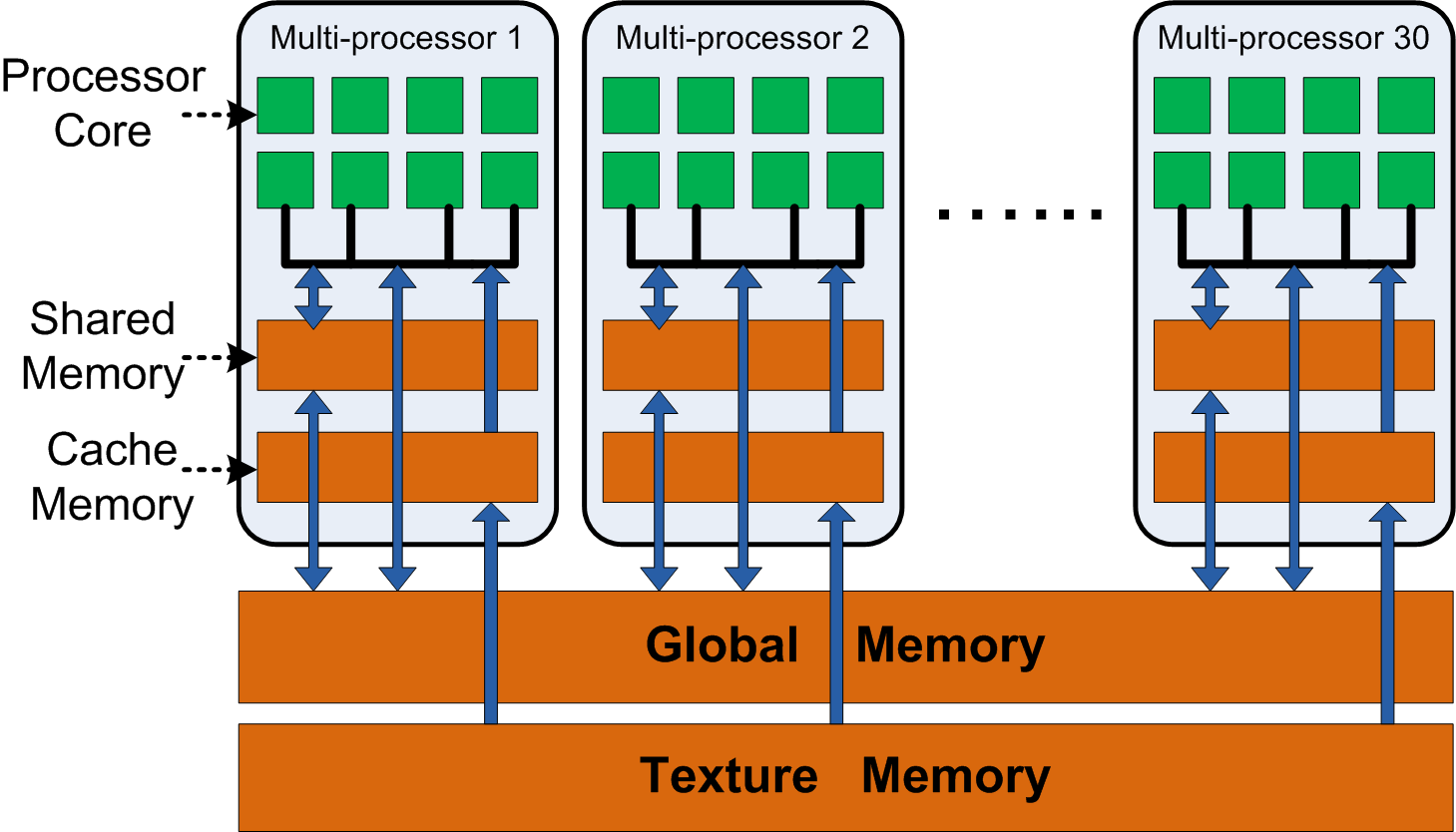}
\caption{The NVIDIA GTX280 GPU architecture.}
\vspace{-0.1in}
\label{fig:gpu}
\end{figure}

%\subsection{Processing Elements}
\noindent
{\bf Processing Elements:}
The basic execution unit on the GTX 280 is a scalar processing
\textbf{core}, of which 8 together form a
\textbf{multiprocessor}. While the number of multiprocessors and
processor clock frequency depends on the make and model of the GPU,
every multiprocessor in CUDA executes in SIMT (Single Instruction,
Multiple Thread) fashion, which is similar in nature to SIMD (Single
Instruction, Multiple Data) execution. Specifically, a group of 32
threads form a \textbf{warp} and are scheduled to execute concurrently. 
However, when codepaths within a warp diverge, the execution of all threads in a warp becomes serialized.
%must now execute every instruction on \emph{every} thread path,therefore, 
This implies that optimal performance is attained when all 32
threads do \emph{not} branch down different codepaths.
%The execution of a single instruction with 8 cores and a warp size of 32 completes in 4 cycles.

%\subsection{Memory Hierarchy}
\noindent
{\bf Memory Hierarchy:}
The GTX 280 contains multiple forms of memory. The read-write \textbf{global memory} and read-only \textbf{texture memory}
is located off-chip on the graphics card and provides the main source
of storage for the GPU, as shown in Figure \ref{fig:gpu}.
% while simultaneously being accessible from the
%CPU and GPU. 
Each multiprocessor on the GPU contains fast on-chip memory, which includes \textbf{cache memory} and \textbf{shared memory}. 
%---a \textbf{texture cache}, \textbf{constant cache}, and \textbf{shared  memory}. 
The texture cache is
\emph{read-only} memory providing fast access to immutable
data. Shared memory, on the other hand, is user-controlled \emph{read-write} space to
provide each core with fast access to the shared address space within a multiprocessor. 
%Finally, on each core resides a plethora of registers
%such that there exists minimal reliance on local memory resident
%off-chip on the device memory.

%\subsection{Parallelism Abstractions}
\noindent
{\bf Parallelism Abstractions:}
At the highest level, the CUDA programming model requires the
programmer to offload functionality to the GPGPU as a compute
\textbf{kernel}. This kernel is evaluated as a set of \textbf{thread
  blocks} logically arranged in a \textbf{grid} to facilitate
organization. In turn, each block contains a set of \textbf{threads},
which will be executed on the same multiprocessor, with the threads
scheduled in warps, as mentioned above.

\subsection{Data Mining using GPGPUs}
\deb{Requires rewording to say algorithm transformation is key to success.}
Many researchers have harnessed the capabilities of GPGPUs for data mining.
The key to porting a data mining algorithm onto a GPGPU is to, in one sense, ``dumb it down''; 
i.e., conditionals, program branches, and complex decision constraints are not
easily parallelizable on a GPGPU and algorithms using these constraints
will require significant reworking to fit this architecture (temporal episode
mining unfortunately falls in this category). There are many emerging publications
in this area but due to space restrictions, we survey only a few here.
The SIGKDD tutorial by Guha et al.~\cite{tutorial} provides a gentle introduction to the aspects of data mining on GPGPUs through problems like
k-means clustering, reverse nearest neighbor(RNN), discrete wavelet transform(DWT), sorting networks, etc.
In~\cite{dmg2}, a bitmap technique is proposed to support counting and is used
to design GPGPU variants of {\it Apriori} and k-means clustering. This work
also proposes co-processing for itemset mining where the complicated tie data
structure is kept and updated in the main memory of CPU and only the
itemset counting is executed in parallel on GPU.
A sorting algorithm on GPGPUs with applications to frequency counting and histogram construction
is discussed in~\cite{dmg1} which essentially recreates sorting networks on the GPU.
%Li et al.~\cite{li-cut} present a `cut-and-stitch' algorithm for approximate learning
%of Kalman filters. Although this is not a GPGPU solution {\it per se}, we point out
%that our proposed approach shares with this work the difficulties of mining temporal behavior 
%in a parallel context. 
%

%# Temporal data mining
%    * Definitions: event stream, episode, occurrence of an episode, non-overlapped count.
%    * The original counting algorithm.

\subsection{Temporal Data Mining}
\label{sec:tdm}
In event sequences, the notion of frequent episodes is used to express patterns of the form $\textrm{A} \rightarrow \textrm{B} \rightarrow \textrm{C}$, i.e., event A is followed (not necessarily consecutively) by B and B is, similarly, followed by C. It is important to constrain the mining by imposing minimum and maximum delays between consecutive symbols in an episode, e.g., to look for episodes of the form $(A^{\underrightarrow{(2,5]}}B^{\underrightarrow{(0,6]}}C)$. This specifies that event A is followed by B within two to five milliseconds, and C follows B within at most six milliseconds. In either unconstrained or constrained episode mining, the occurrences of an episode allow `junk' or don't care events, of arbitrary length, between the event symbols of the episode. This is what makes these patterns very useful and comprehensible. Many frequency measures~\cite{vatsan1,window} for episodes have been defined that obey anti-monotonicity and hence search for such episodes can be structured level-wise, ala {\it Apriori}. The first measure to be proposed was the window based frequency measure \cite{window}. Later the notion of \textit{non-overlapped} occurrences count was shown to have properties that enable fast sequential counting algorithms \cite{vatsan1}.

%\vspace{0.1in}

\begin{definition}
An \textit{Event Stream} is denoted by a sequence of events $\langle(E_1, t_1),(E_2, t_2),\ldots(E_n, t_n)\rangle$, where $n$ is the total number of events. Each event $(E_i, t_i)$ is characterized by an event type $E_i$ and a time of occurrence $t_i$. The sequence is ordered by time i.e. $t_i\leq t_{i+1} \forall i = 1,\ldots,n-1$ and $E_i$'s are drawn from a finite set $\xi$.
\end{definition}

In neuroscience, a spike train is a series of discrete action potentials from several neurons generated spontaneously or as a response to some external stimulus. This data neatly fits into the frequent episodes framework of analyzing event streams.

%\vspace{0.1in}

\begin{definition}
An (serial) episode $\alpha$ is an ordered tuple of event types $V_{\alpha} \subseteq \xi$.
\end{definition}
For example $(A \rightarrow B \rightarrow C)$ is a 3-node serial episode, and it captures the pattern that neuron A fires, followed by neurons B, and C in that order, but not necessarily consecutive.

%\vspace{0.1in}

\textit{Frequency of episodes}: 
A frequent episode is one whose frequency exceeds a user specified threshold. 
The notion of frequency of an episode is intended to capture the repeating nature of an episode in an event sequence.
In this work we shall focus on the measure of frequency defined as the size of the largest set of \textit{non-overlapped} occurrences of an episode \cite{vatsan1}.

%\vspace{0.1in}

\textit{Temporal constraints}: Episodes can be further specialized by specifying constraints on the timing of the events in episode occurrences \cite{Patnaik2008}.
Placing inter-event time constraints giving rise to episodes of the form:
\begin{equation*}
(A^{\underrightarrow{(t_{low}^{1},t_{high}^{1}]}}B^{\underrightarrow{(t_{low}^{2},t_{high}^{2}]}}C)
\label{eq:episode-with-interval}
\end{equation*}

That is, in an occurrence of episode $A\rightarrow$ $B\rightarrow$ $C$ 
let $t_{A}$, $t_{B}$, and $t_{C}$ denote the time of occurrence of corresponding event types. A valid occurrence of the serial episode satisfies $$t_{low}^{1} < (t_{B}-t_{A}) \le t_{high}^{1},
t_{low}^{2} < (t_{C}-t_{B}) \le t_{high}^{2}.$$
(In general, an $N$-node serial episode is associated with $N-1$ inter-event constraints.)

Level-wise discovery procedure for frequent episodes finds the complete set of episodes with frequency or count greater than a user-defined threshold. At each level ($N$)-size candidates are generated from ($N-1$)-size frequent episodes and their count is determined by making a pass over the event sequence. Only those candidates whose count is greater than the threshold are retained. The event sequences typically runs very long and hence the counting step is computationally most expensive. For initial passes of the algorithm where we have several short episodes to count the counting task can be embarrassingly parallel by making each thread of execution count occurrences of only one episode. But in later stages there are relatively fewer but longer episodes leading to severe under-utilization. We focus our work here on the later situation and attempt to solve this counting problem.

\section{Algorithm}
\subsection{Episode Counting Algorithm}

\begin{figure}[htbp]
	\centering
		\includegraphics[width=0.8\columnwidth]{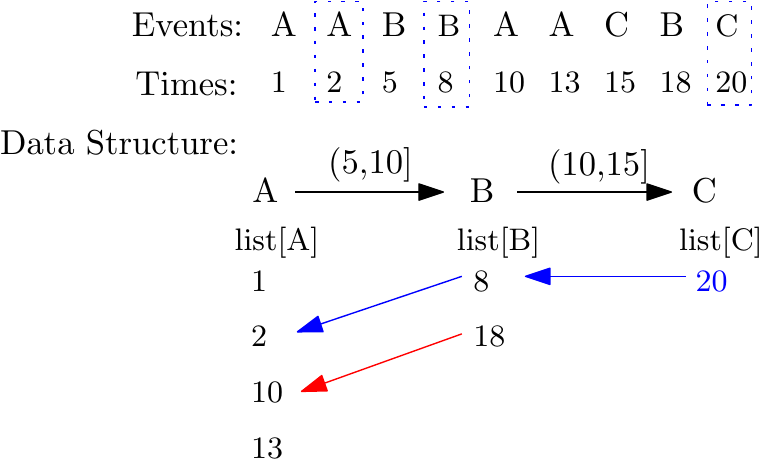}
	\caption{Illustration of the algorithm for counting non-overlapped occurrences of $A\stackrel{(5,10]}{\rightarrow}B\stackrel{(10,15]}{\rightarrow}C$. This proceeds from left-to-right of the event steam keeping track of sufficient information required to obtain the correct non-overlapped count under the given inter-event time constraints.}
	\label{fig:counter-eg}
\vspace{-0.1in}
\end{figure}

The algorithm presented in~\cite{Patnaik2008} is based on finite state machines. This sequential algorithm for mining frequent episodes with inter-event constraints works by maintaining a data-structure shown in Figure~\ref{fig:counter-eg} as the read head moves down the event sequence. In this example we are counting occurrences of episode 
$A\stackrel{(5,10]}{\rightarrow}B\stackrel{(10,15]}{\rightarrow}C$.

%\begin{algorithm}[htpb]
%\begin{algorithmic}[1]
%\REQUIRE Candidate $N$-node episode $\alpha=\langle E_{(1)}\stackrel{(t_{low}^{(1)},t_{high}^{(1)}]}{\longrightarrow}\ldots E_{(N)}\rangle$ and event sequence $\textsl{S}=\{(E_i, t_i)\}, i \in \{1\ldots{n}\}$.
%\ENSURE Count of non-overlapped occurrences of $\alpha$ satisfying inter-event constraints
%\STATE $count = 0$; $s=[[],\ldots,[]]$ //List of $|\alpha|$ lists
%\FORALL{$(E,t) \in \textsl{S}$}
%	\FOR{$i = |\alpha| \mbox{ to } 1$} \label{line:outer}
%		\STATE $E_{(i)} = i^{th}$ event type $\in \alpha$
%		\IF{$E = E_{(i)}$}
%			\STATE $i_{prev} = i - 1$
%			\IF{$i > 1$}
%					\STATE /* Remove old entires in $s[i_{prev}]$*/
%					\WHILE{$|s[i_{prev}]|>0$ and $s[i_{prev},0] < t - t_{high}^{i_{prev}}$}
%						\STATE delete $s[i_{prev},0]$
%					\ENDWHILE
%				\STATE $k = |s[i_{prev}]|$
%				\WHILE{$k > 0$}\label{line:inner}
%					\STATE $t_{prev} = s[i_{prev},k]$
%					\IF{$t_{low}^{(i_{prev})} < t - t_{prev} \leq t_{high}^{(i_{prev})}$}
%						\IF{$i = |\alpha|-1$}
%							\STATE $count ++$; $s = []$; \textbf{break} Line: \ref{line:outer}
%						\ELSE
%							\STATE $s[i]=s[i]\cup{t}$
%						\ENDIF
%						\STATE \textbf{break} Line: \ref{line:inner}
%					\ENDIF
%					\STATE $k = k - 1$
%				\ENDWHILE
%			\ELSE % i_{prev} == 0
%				\STATE $s[i]=s[i]\cup{t}$
%			\ENDIF
%		\ENDIF
%	\ENDFOR
%\ENDFOR
%\STATE RETURN count
%\end{algorithmic}
%\caption{Serial Episode Mining}
%\label{alg:A1}
%\end{algorithm}

The general approach, on finding an event that belongs to the episode, is to look up the list of occurrences of the previous symbol (or event-type). If there exists an event of the previous event type which together with the new event satisfies the inter-event constraint for the pair of event-types, then the new event is added to the data-structure under its corresponding symbol. An occurrence is said to be complete when we can add an event for the last symbol in the episode. Then the count is incremented and the data-structure cleared. For example when $(B,18)$ arrives, it is found that the pair $(A,10)\in list[A]$ and $(B,18)$ satisfy the inter-event constraint $(5,10]$. And therefore the time for $(B,18)$ is recoded in $list[B]$. On arrival of $(C,20)$ we are able to complete one occurrence of the episode.

%For completeness the counting procedure is reproduced in Algorithm~\ref{alg:A1} from \cite{Patnaik2008}.

\subsection{MapConcat}
The initial idea for parallelization is to map one thread per episode. However, when the number of episodes is smaller than the number of cores, a one-thread per-episode model is prone to under utilization leading to higher execution times. Hence our first attempt to achieve a higher level of parallelism within the counting of a single episode was to segment the input event stream into several of sub-streams and use one thread block to count one episode \cite{Yong2009}.

\begin{figure}[htpb]
	\centering
		\includegraphics[width=0.8\columnwidth]{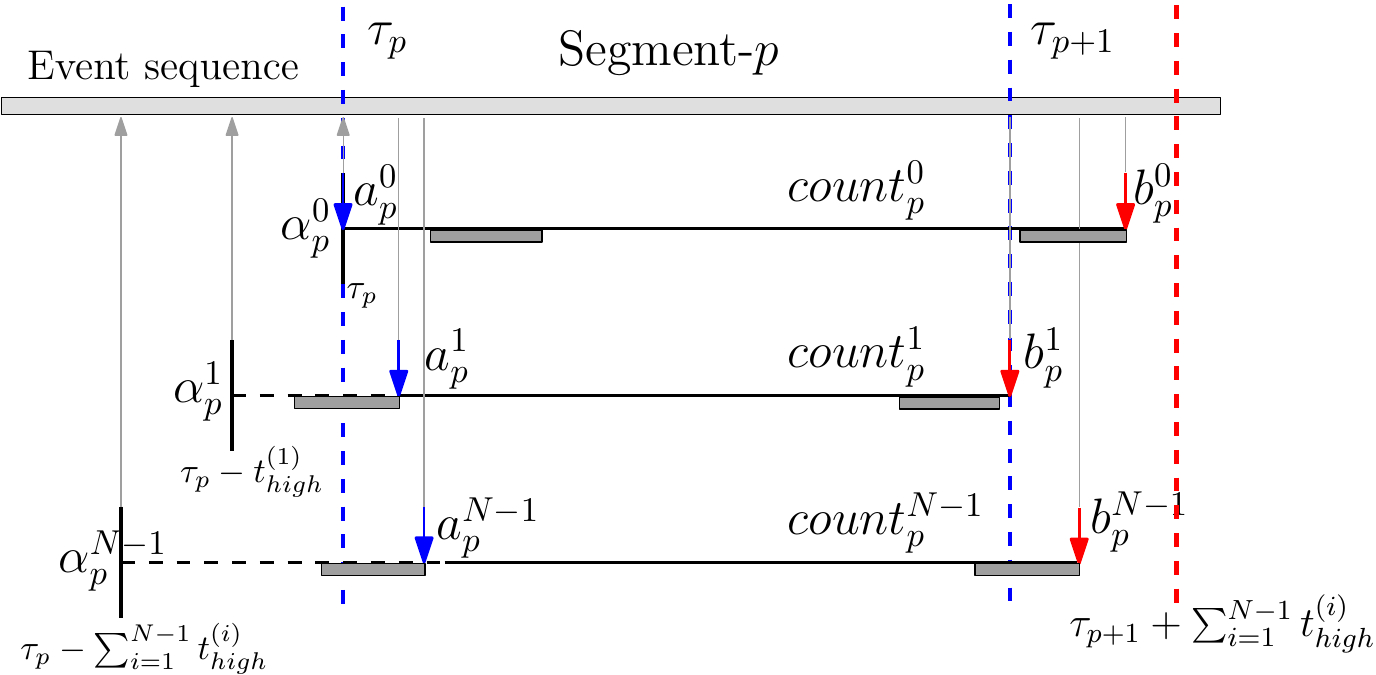}
	\caption{Illustration of a {\em Map} step. Multiple state machines are needed to track occurrences of $N$-size episode $\alpha$ in the $p^{th}$-Segment of the data. Each state machine starts at a different offset into the previous segment and continues over into the next segment to complete the last occurrence. The end-time of the first $a$ and last occurrence $b$ seen by a state machine are recorded beside the $count$ for the reduce/concat step.}
	\label{fig:map}
\end{figure}

\begin{figure}[htpb]
	\centering
		\includegraphics[width=0.8\columnwidth]{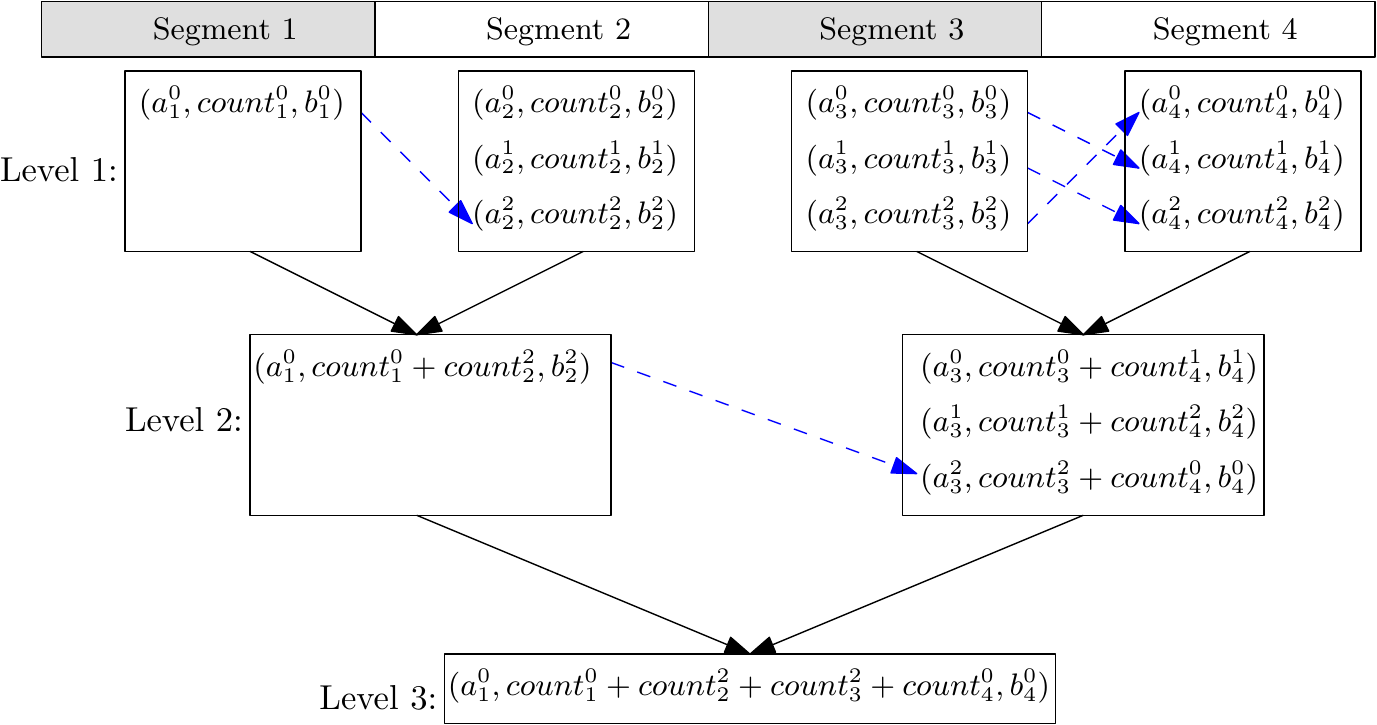}
	\caption{Illustration of a {\em Concatenate} step. As noted in Fig~\ref{fig:map}, the first and last occurrence of each state-machine is used to stitch together the total count for an episode. The blue arrow in this figure indicates that the last occurrence seen by the state-machine on the left matches with the first occurrence seen by the state-machine on the right. These state-machines can then be merged together into the next level.}
	\label{fig:concat}
\end{figure}

 When we divide the input stream into segments, there are chances that some occurrences of an episode span across the boundaries of consecutive segments. It turns out that in order to obtain the correct count we are required to run multiple state machines within the same data segment anticipating all possible end states of the state machine in the previous segment. The final count is obtained by a reduce step where state machines for consecutive segments with matching start and end states are concatenated. The counting step (or the Map-step) is illustrated in Fig.~\ref{fig:map} and the reduce step (or Concat-step) is shown in Fig.~\ref{fig:concat}.

%% file: tex/priorwork.tex
%\section{Prior Work}

%% file: tex/algorithms.tex
\section{Redesigned Algorithm}
 \yong{Motivation:\\

    * Why the approach of hardware-based optimization is limited?\\
    * Introduce the concept of AOAT. Better with one or two examples.\\
    * In temporal data mining, the original algorithm is state-machine based\\ counting which has lot of data dependencies (non data parallel). It's very\\ difficult to increase the degree of parallelism by optimizing the original algorithm.
    
Redesigned algorithm\\

    * Problem decomposition\\
          o State two sub-problems (the solution for the second problem is trivial and it's not computational expensive )\\
    * Parallel local tracking (for the first problem)\\
          o Introduce parallel local tracking (with figures and pseudo code)\\
          o There are two steps: tracking and compacting. Describe the GPU\\ implementation for tracking (embarrassingly parallel). State of difficulty of GPU compacting (non data parallel).\\
          o Propose two solutions for GPU compacting (atomic function and pre-fix scan). \\
                + Compare the difference (atomic function has fast kernel execution but requires sorting. Pre-fix scan does not require sorting. \\
                + For prefix scan, state the limitation of CUDPP. Introduce our implementation.}

The original state-machine based algorithm has a lot of data dependency. It is difficult to increase the degree of parallelism by optimizing this algorithm. In order to transform the algorithm to map well onto the GPGPU architecture, we must revisit the problem and
formulate a decomposition using known parallel primitives.
                
\subsection{Problem decomposition}
Referring to the definition of non-overlapped frequency of an episode, it is the size of the largest set of non-overlapped occurrences. Based on this definition we design a more data parallel solution. In our new approach we find a super-set of the non-overlapped occurrences which could potentially be overlapping. Each occurrence of an episode has fixed start and end times. If each episode occurrence in this super set can be viewed as a task with start and end time, then the problem of finding the largest set of non-overlapped occurrences can be easily transformed into a task scheduling problem, where the goal is to maximize number of non-conflicting tasks on a single shared resource. It is known that a greedy $O(n)$ algorithm can solve this problem optimally. The original problem now decomposes into the following two sub-problems:
\begin{itemize}
	\item \textit{Subproblem-1}: Finding a super-set of non-overlapped occurrences of an episode.
	\item \textit{Subproblem-2}: Finding the size of the largest set of non-overlapped occurrences from the above set of occurrences.
\end{itemize}

The first sub-problem can be solved with high parallelism as will be shown in following sections. The second sub-problem is same as the task or interval scheduling problem where tasks have fixed times. A fast greedy algorithm is known to solve the problem optimally.

We first pre-process the entire event stream noting the positions of events of each event-type. Then for a given episode, beginning with the list of occurrences of the start event-type in the episode, we find occurrences satisfying the temporal constraints in parallel. Finally we collect and remove overlapped occurrences in one pass. The greedy algorithm for removing overlaps requires the occurrences to be sorted by end time and the algorithm proceeds as shown in Algorithm~\ref{alg:sch}. In this algorithm, for every set of
consecutive occurrences, if the start time is after the end time of the
last selected occurrence then we select this occurrence, otherwise we skip it and go to the next occurrence.

\begin{algorithm}[htpb]
\caption{Obtaining the largest set of non-overlapped occurrences}
\label{alg:sch}
\begin{algorithmic}
\REQUIRE{List $\mathcal{C}$ of occurrences with start and end times $(s_i, e_i)$ sorted by end time, $e_i$.}
\ENSURE{Size of the largest set of non-overlapped occurrences}
\STATE Initialize $count=0$
\STATE $prev_e = 0$
\FORALL{$(s_i, e_i) \in \mathcal{C}$}
	\IF{$prev_e < s_i$}
%		\STATE \COMMENT{Select the current task}
		\STATE $prev_e = e_i$; $count = count + 1$
	\ENDIF
\ENDFOR
\STATE return $count$
\end{algorithmic}
\vspace{-0.02in}
\end{algorithm}

\vspace{-0.1in}
\subsection{Finding occurrences in parallel}
Here we explore different approaches of solving the first sub-problem as discussed in the previous section. The aim here is to find a super-set of non-overlapped occurrences in parallel. The basic idea is to start with all events of the first event-type in parallel for a given episode and find occurrences of the episode starting at each of these events. There can be several different ways in which this can be done. We shall present two approaches that gave us the most performance improvement. We shall use the episode $A\stackrel{(5-10]}{\longrightarrow}B\stackrel{(5-10]}{\longrightarrow}C$ as our running example and
explain each of the counting strategies using this example. This example episode specifies event occurrences where an event A is to be
followed by an event B within 5-10 ms and event B is to be
followed by an event C within 5-10 ms delay. Note again that the delays have both a lower and a upper bound.

\subsection{Parallel Local Tracking}
\begin{figure}%
\centering
\includegraphics[width=0.8\columnwidth]{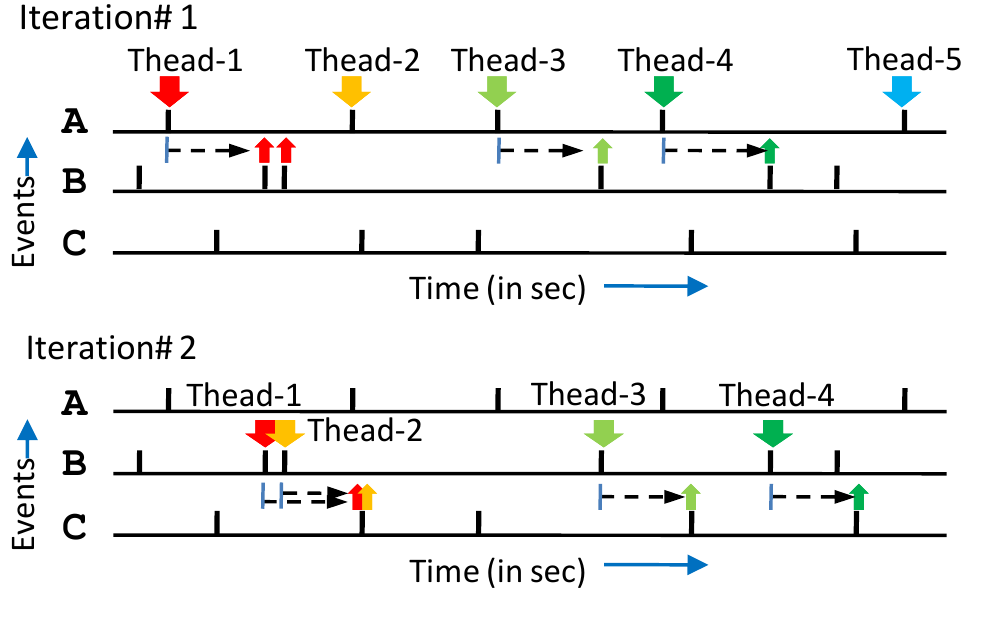}%
\vspace{-0.1in}
\caption{Illustration of Parallel local tracking algorithm (See Algorithm~\ref{alg:kernel}), showing 2 iterations for the episode $A \rightarrow B \rightarrow C$ with implicit inter-event constraints. Note that each thread can find multiple next-events. Further, a thread stops scanning the event sequence when event-times go past the upper bound of the inter-event constraint.}%
\vspace{-0.1in}
\label{fig:kernel}%
\end{figure}

In the pre-processing step, we have noted the locations of each of the event-types in the data. In the counting step, we launch as many threads as there are events of the start type. In our running example this is all events of type $A$. Each thread searches the event streams starting at one of these events of type $A$ and looks for an event of type $B$ that satisfies the inter-event time constraint $(5-10]$ i.e., $5 < t_{B_j} - t_{A_i}\leq 10$ where $i,j$ are the indices of the events of type $A$ and $B$. One thread can find multiple $B$'s for the same $A$. These are recorded in a preallocated array assigned to each thread. Once all the events of type $B$ (with an $A$ before them) have been collected by the threads (in parallel), we need to compact these newly found events into a contiguous array/list. This is necessary as in the next kernel launch we will find all the events of type $C$ that satisfy the inter-event constraints with this set of $B$'s. This is illustrated in Figure~\ref{fig:kernel}.

\begin{algorithm}[htpb]
\caption{Kernel for Parallel Local Tracking}
\label{alg:kernel}
\begin{algorithmic}
	\REQUIRE Iteration number $i$, Episode $\alpha$, Event type $\alpha[i]$, Index list $I_{\alpha[i]}$, Data sequence $S$.
	\ENSURE Indices of events of type $\alpha[i+1]$.
	\FORALL{Threads which have distinct identifiers $tid$}
		\STATE Scan $S$ starting at event $I_{\alpha[i]}[tid]$ for event-type $\alpha[i+1]$ satisfying inter-event constraint $(t_{low}^{(i)},t_{high}^{(i)}]$.
		\STATE Record all such events of type $\alpha[i+1]$.
	\ENDFOR
	\STATE Compact all found events into the list $I_{\alpha[i+1]}$.
	\RETURN $I_{\alpha[i+1]}$
\end{algorithmic}
\vspace{-0.045in}
\end{algorithm}

Algorithm~\ref{alg:kernel} presents the over-all work done in each kernel launch. In order to obtain the complete set of occurrences of an episode, we need to launch the kernel $N-1$ times where $N$ is the size of an episode. The list of events found in the $i^{th}$ iteration is passed as input to the next iteration. Some amount of book-keeping is also done to keep track of the start and end times of an occurrence. After this phase of parallel local tracking is completed the non-overlapped count is obtained using Algorithm~\ref{alg:sch}.
The compaction step in Algorithm~\ref{alg:kernel} presents a challenge as it requires concurrent updates into a global array.

\subsection{Lock-based compaction}
\begin{figure}%
\centering
\includegraphics[width=0.8\columnwidth]{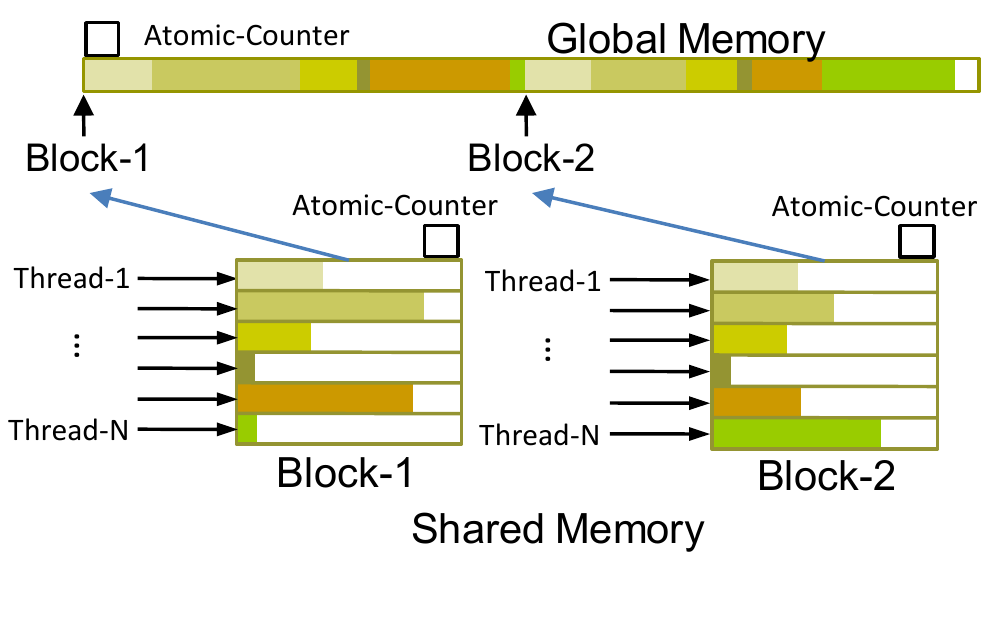}%
\vspace{-0.2in}
\caption{Illustration of output compaction using AtomicAdd operations. Note that we use atomic operations at both block and global level. These operations return the correct offset into global memory for each thread to write its next-event list into.}%
\vspace{-0.05in}
\label{fig:atomic}%
\end{figure}

Nvidia graphics cards with CUDA compute capability 1.3 support atomic operations on shared and global memory. 
Here we use atomic operations to perform compaction of the output array into the global memory. After the counting step each thread has a list of next-events. Subsequently each thread adds the size of its next-events list to the block-level counter using an atomic add operation and in return obtains a local offset (which is the previous value of the block-level counter). After all threads in a block have updated the block-level counter, the first thread in the block updates the global-counter by adding the value of the block-level counter to it and,
as before, obtains the offset into global memory. Now all threads in the block can collaboratively write into the correct position in the global memory (resulting in overall compaction). A schematic for this operation is shown for 2-blocks in Figure~\ref{fig:atomic}. In the results section, we refer to this method as \textit{AtomicCompact}.

Since there is no guarantee for the order of atomic operations, this procedure requires sorting. The complete occurrences need to be sorted by end time for Algorithm~\ref{alg:sch} to produce the correct result.
%We do this sorting on the CPU as sorting on the GPU is not found to be as efficient.

\subsection{Lock-free compaction}

Prefix-scan is known to be a general-purpose data-parallel primitive that is a useful building block for algorithms in a broad range of applications. Given a vector of data elements $[x_0,x_1,x_2,\ldots]$, an associative binary function $\oplus$ and an identity element $i$, exclusive prefix-scan returns $[i,x_0, x_0\oplus{x_1}, x_0\oplus{x_1}\oplus{x_2},\ldots]$.  Although the problem is seemingly sequential the first parallel prefix-scan algorithm was proposed in 1962 \cite{Iverson1962}. With recently increasing interest is GPGPU, several implementations of scan have been proposed for GPU, the most recent being \cite{Harris2007, Sengupta2007}. This later implementation is available as the \textit{CUDPP: CUDA Data Parallel Primitives Library} and forms part of the CUDA SDK distribution. 

Our lock-free compaction is based on prefix-sum and we reuse the implementation from CUDPP library. Since the scan based operation guarantees ordering we modify our counting procedure to count occurrences backwards starting from the last event. This results in the final set of occurrences to be automatically ordered by end-time and therefore completely eliminates the need for sorting (as required by atomic operations based approach).

%\deb{TODO: Scan description. Our implementation desc.}

\begin{figure}[ht]
\centering
\includegraphics[width=0.8\columnwidth]{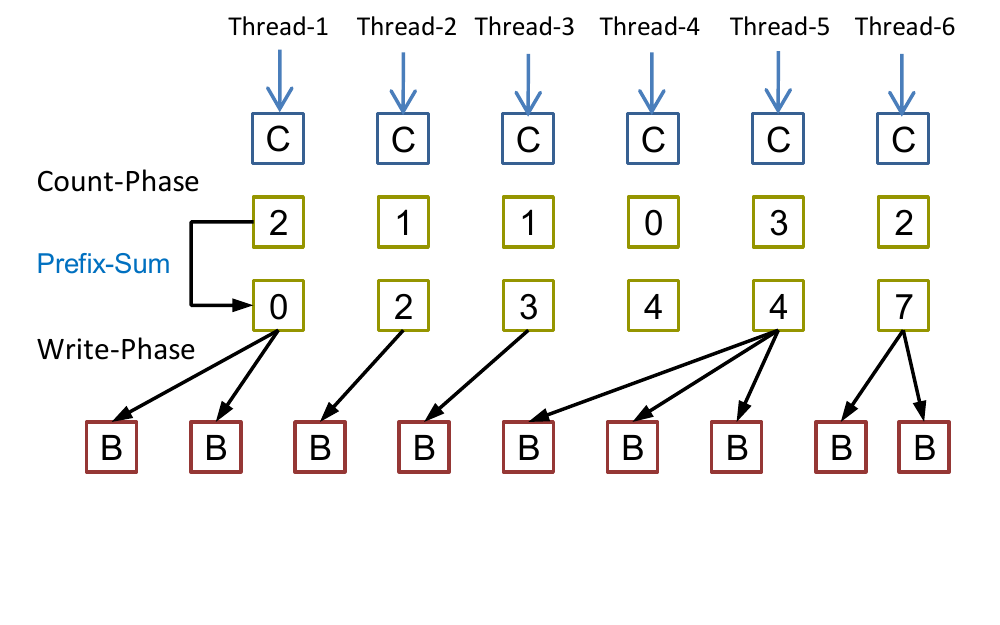}%
\vspace{-0.4in}
\caption{Illustration of output compaction using Scan primitive. Each iteration is broken into 3 kernel calls: Counting the number of next events, Using scan to compute offset into global memory, finally launching count-procedure again but this time allowing write operations to the global memory}%
\label{fig:scan}%
\vspace{-0.05in}
\end{figure}

%cudppCompact
The CUDPP library provides a compact function which takes an array $d_{in}$, an array of $1/0$ flags and returns a compacted array $d_{out}$ of corresponding only the ``valid'' values from $d_{in}$ (it internally uses \texttt{cudppScan}). In order to use this, our counting kernel is
now split into three kernel calls. Each thread is allocated a fixed portion of a larger array in global memory for its next-events list. In the first kernel, each thread finds its events and fills up its next-events list in global memory. The \texttt{cudppCompact} function (implemented as two GPU kernel calls) compacts the large array to obtain the global list of next-events. 
%This approach has several difficulties, firstly, 
A difficulty of this approach is that the array on which \texttt{cudppCompact} operates is very large resulting in a scattered memory access pattern.
%, and secondly shared memory is not utilized at all.
We refer to this method as \textit{CudppCompact}.

%Count-Scan-Write
In order to improve performance, we adopt a counter-intuitive approach. We divide the counting process into three parts. First each thread looks up the
event sequence for suitable next-events but instead of writing the found events anywhere, it merely counts and writes the count to global memory. Then an exclusive scan is performed for the count-array which gives the offset into the global memory where each thread can write its next-events list. The actual writing is done as the third step. Although each thread looks up the event sequence twice (first to count, and second to write) we show that we nevertheless
obtain better performance.
This entire procedure is illustrated in Figure~\ref{fig:scan}. We refer to this method of compaction as \textit{CountScanWrite} in the ensuing results section.

%Blockwise-compaction followed by global compaction
%We further try to improve the above approach by performing compaction in two levels first within a block (using shared memory) followed by compaction in global memory. But this performs only slightly better than the previous approach.

%% file: tex/results.tex
\section{Results}
The hardware used for obtaining the performance results 
are given in Table~\ref{tab:hardware}:

\begin{table}[htbp]
\centering
\caption{Hardware used for performance analysis}
\label{tab:hardware}
\begin{tabular}{|l|c|}
\hline
GPU & Nvidia GTX 280 \\ \hline
Memory (MB) & 1024 \\
Memory Bandwidth (GBps) & 141.7\\
Multiprocessors, Cores & 30, 240 \\
Processor Clock (GHz)& 1.3 \\ \hline
CPU & Intel Core 2 Quad Q8200\\ \hline
Processors & 4\\
Processor Clock (GHz) & 2.33\\
Memory (MB) & 4096\\ \hline
\end{tabular}
\end{table}

\subsection{Test datasets and algorithm implementations}
The datasets used here are generated from the non-homogeneous Poisson process model for inter-connected neurons described in \cite{Patnaik2008}. This simulation model generates fairly realistic spike train data. For the datasets in this paper a networks of 64 artificial neurons was used. The random firing rate of each neuron was set at 20 spikes/sec to generate sufficient noise in the data. Four 9-node episodes were embedded into the network by suitably increasing the connection strengths for pairs of neurons. Spike train data was generated by running the simulation model for different durations of time. Table~\ref{tab:datasets} gives the duration and number of events in each dataset.

\begin{table}[htpb]
\caption{Details of the datasets used}
\label{tab:datasets}
\centering
\begin{tabular}{cc}
{
\begin{tabular}{|c|c|c|}
\hline
Data & Length & \# Events \\
-Set & (in sec) & \\ \hline
1 & 4000 & 12,840,684 \\
2 & 2000 & 6,422,449 \\
3 & 1000 & 3,277,130 \\
4 & 500  & 1,636,463 \\ \hline
\end{tabular}
} & \hspace{-0.2in}
{
\begin{tabular}{|c|c|c|}
\hline
Data & Length & \# Events \\
-Set & (in sec) & \\ \hline
5 & 200  & 655,133   \\
6 & 100  & 328,067   \\
7 & 50   & 163,849   \\
8 & 20   & 65,428    \\ \hline
\end{tabular}
}
\end{tabular}
\end{table}

In Section \ref{sec:perf-compare}, we compare the performance of our new algorithm to \textit{MapConcat} and a CPU implementation of the original algorithm described in Section \ref{sec:tdm}. In order to analyze the lock-based and lock-free compaction strategies, we present the performance of a lock-based method, \textit{AtomicCompact}, and two lock-free methods, \textit{CudppCompact} and \textit{CountScanWrite}, as shown in Section \ref{sec:new-alg-analysis}.

\subsection{Comparisons of performance}
\label{sec:perf-compare}

\begin{figure}[htbp]
\centering
\includegraphics[width=0.85\columnwidth]{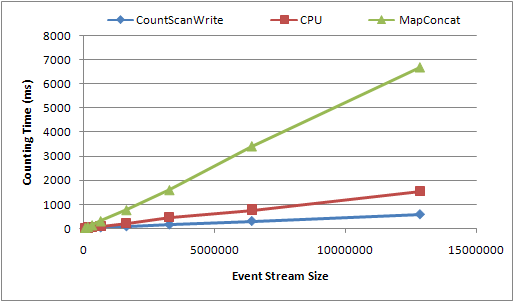}
\caption{Performance of MapConcat compared with the CPU and best GPU implementation, counting 30 episodes in Datasets 1-8.}
\vspace{-0.03in}
\label{fig:mapconcat-performance}
\end{figure}

\textit{MapConcat} is clearly a poor performer compared to the CPU, with up to a 4x slowdown. Compared to our best GPU method, \textit{MapConcat} is up to 11x slower. This is due to the overhead induced by the merge step of \textit{MapConcat}. Although at the multiprocessor level each episode is counted in parallel, the logic required to obtain the correct count is complex.

We run the CUDA Visual Profiler on \textit{MapConcat} and one of our redesigned algorithms, \textit{CountScanWrite}. Dataset 2 was used for profiling each implementation. Due to its complexity, \textit{MapConcat} exhibited poor features such as large amounts of divergent branching
%, local loads and stores, 
and a large total number of instructions executed, as shown in Table \ref{tab:profiler}. Comparatively, the \textit{CountScanWrite} implementation only exhibits divergent branching.%, with no local loads or stores. 

\begin{table}[htbp]
\centering
\caption{CUDA Visual Profiler Results}
\label{tab:profiler}
\begin{tabular}{|l|c|c|}
\hline
 & MapConcat & CountScanWrite \\ \hline
Instructions & 93,974,100 & 8,939,786 \\
Branching & 27,883,000 & 2,154,806 \\
Divergent Branching & 1,301,840 & 518,521 \\ \hline
%Local Load & 7,336,750 & 0 \\
%Local Store & 14,317,700 & 0 \\ \hline
\end{tabular}
\end{table}

\begin{figure}[htbp]
\centering
\includegraphics[width=0.85\columnwidth]{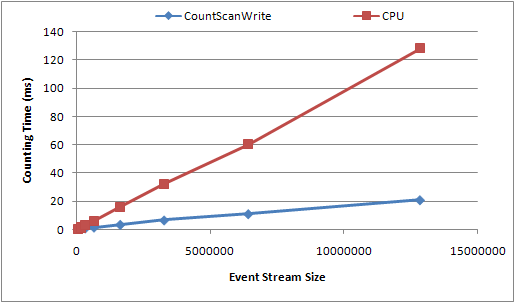}
\caption{Performance comparison of the CPU and best GPU implementation, counting a single episode in Datasets 1 through 8.}
\vspace{-0.03in}
\label{fig:cpu-gpu-performance}
\end{figure}

In terms of the performance of our best GPU method, we achieve a 6x speedup over the CPU implementation on the largest dataset, as shown in Figure \ref{fig:cpu-gpu-performance}.
%At the largest datasets, the best GPU implementation achieves a 6x speedup over the CPU implementation.

\subsection{Analysis of the new algorithm}
\label{sec:new-alg-analysis}

\begin{figure}[htbp]
\centering
\includegraphics[width=0.85\columnwidth]{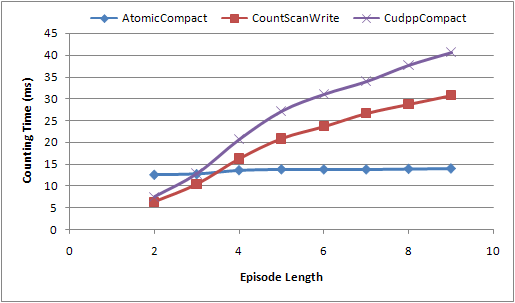}
\caption{Performance of algorithms with varying episode length in Dataset 1.}
\vspace{-0.03in}
\label{fig:length-vs-performance}
\end{figure}

Figure \ref{fig:length-vs-performance} contains the timing information of three compaction methods of our redesigned GPU algorithm with varying episode length. Compaction using CUDPP is the slowest of the GPU implementations, due to its method of compaction. It requires each data element to be either in or out of the final compaction, and does not allow for compaction of groups of elements. For small episode lengths, the \textit{CountScanWrite} approach is best because sorting can be completely avoided. However, at longer episode lengths, compaction using lock-based operators shows the best performance. This method of compaction avoids the need to perform a scan and a write at each iteration, at the cost of sorting the elements at the end. The execution time of the \textit{AtomicCompact} is nearly unaffected by episode length, which seems counter-intuitive because each level requires a kernel launch. However, each iteration also decreases the total number of episodes to sort and schedule at the end of the algorithm. Therefore, the cost of extra kernel invocations is offset by the final number of potential episodes to sort and schedule.

\begin{figure}[htbp]
\centering
\includegraphics[width=0.85\columnwidth]{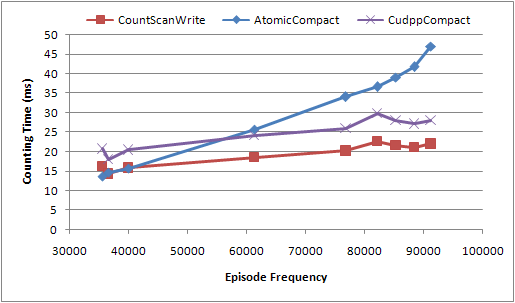}
\caption{Performance of algorithms with varying episode frequency in Dataset 1.}
\vspace{-0.03in}
\label{fig:frequency-vs-performance}
\end{figure}

We find that counting time is related to episode frequency as shown in Figure \ref{fig:frequency-vs-performance}. There is a linear trend, with episodes of higher frequency require more counting time. The lock-free compaction methods follow an expected trend of slowly increasing running time because there are more potential episodes to track. The method that exhibits an odd trend is the lock-based compaction, \textit{AtomicCompact}. As the frequency of the episode increases, there are more potential episodes to sort and schedule. The running time of the method becomes dominated by the sorting time as the episode frequency increases.

Another feature of Figure \ref{fig:frequency-vs-performance} that requires explanation is the bump where the episode frequecy is slightly greater than 80,000. This is because it is not the final non-overlapped count that affects the running time, it is the total number of overlapped episodes found before the scheduling algorithm is appiled to remove overlaps. The x-axis is displaying non-overlapped episode frequency, where the run-time is actually affected more by the overlapped episode frequency.

We used the CUDA Visual Profiler on the other GPU methods. They had similar profiler results as the \textit{CountScanWrite} method. The reason is that the only bad behavior exhibited by the method is divergent branching, which comes from the tracking step. This tracking step is common to all of the GPU method of the redesigned algorithm.

%% file: tex/discussion.tex
\section{Conclusion}
Just as algorithms for secondary storage are quite distinct from algorithms
for main memory (e.g., two-phase merge sort is preferable over quicksort),
we have shown similarly that approaches for temporal data mining on a GPU must
adopt fundamentally different strategies than that on a CPU. Through this
work, we aim to have conveyed that even with an application such as frequent
episode mining that is uneqivocally `sequential' in nature, it is possible to
obtain reasonable speedup by an order of magnitude using careful redesign
and algorithm-oriented transformation. A key lesson from our efforts is
the importance of investigating both ``in--the--large'' and ``in-the-small''
issues. We have demonstrated the effectiveness of our approach to handling
large scale event stream datasets modeled by inhomogeneous Poisson processes.

\section{Future Work}
Similar to our motivations stemming from computational neuroscience, there are
a large class of data mining tasks in bioinformatics, linguistics, and
event stream analysis that require analysis of sequential data.
Our work opens up the interesting issue of the extent to which finite
state-machine based algorithms for these tasks
can be accelerated using GPU platforms. 
Are there fundamental limitations to porting such algorithms on GPUs?
We believe there are
and hope to
develop a theoretical framework to investigate GPU-transformation
issues for these algorithms. Second, we aim to study the development of
streaming versions of these algorithms where approximate results are
acceptable but near real-time responsiveness is important.